# SLEAN: Simple Lightweight Ensemble Analysis Network for Multi-Provider LLM Coordination: Design, Implementation, and Vibe Coding Bug Investigation Case Study

Matheus J. T. Vargas*

*Affiliation: CyAura AI, matheus@cyaura.ai

**ABSTRACT**

We present SLEAN (Simple Lightweight Ensemble Analysis Network), a deterministic framework for coordinating multiple LLM providers through text-based prompt orchestration. Unlike complex multi-agent systems requiring specialized infrastructure, SLEAN operates as a simple prompt bridge between LLMs using .txt templates, requiring no deep technical knowledge for deployment. The three-phase protocol formed by independent analysis, cross-critique, and arbitration, filters harmful AI-generated code suggestions before production deployment, addressing how AI-assisted debugging increasingly produces modifications that introduce unnecessary complexity, break existing functionality, or address problems. Evaluating 15 software bugs, we analyzed 69 AI-generated fix propositions. SLEAN's filtering accepted 22 fixes (31.9%, 95% CI 20.9-42.9%) while rejecting 47 that would have been harmful if applied verbatim. The arbitration process reduced code change surface by 83-90% relative to raw AI outputs, enforcing minimal causal edits over scope-expanding modifications. Minimal Type 2 inputs proved more efficient than detailed Type 1 inputs, requiring 2.85 versus 3.56 propositions per accepted fix (35.1% versus 28.1% acceptance, about a 20% efficiency gain). Agreement between AI systems showed weak correlation with fix quality: high convergence (at least 80%) occurred in 4 of 15 cases and improved acceptance by only 2.4% points; arbitration appeared only at exactly 10% convergence in 2 of 15 cases, although low convergence alone did not necessitate arbitration. The file-driven, provider-agnostic architecture enables deployment without specialized coding expertise, making it applicable to security auditing, code review, document verification, and other domains requiring reliable multi-provider synthesis with end-to-end auditability.

**BACKGROUND**

Large Language Models (LLMs) have shown significant promise in data analysis, code generation, and automated program repair, yet practical deployment faces a fundamental tension between analytical sophistication and deployment complexity.[1–4] Single-model approaches suffer from inherent biases and inconsistencies that can yield incomplete or incorrect answers, especially in complex problems.[5] Meanwhile, sophisticated multi-agent frameworks demonstrate impressive collaborative reasoning but require complex, stateful runtime environments that are impractical for many production pipelines as well as for general use, for example, vibe coders.[6,7]

Existing orchestration frameworks such as CrewAI, AutoGen, LangGraph, Swarm, and Voyager excel at multi-agent collaboration through persistent agent memory, complex role negotiation, and graph-based orchestration.[5,8–10] However, these capabilities come at substantial infrastructure cost and the use of single choice LLM.[11] Complex problems, such as production debugging workflows, require solutions that integrate seamlessly with existing continuous integration/continuous delivery (CI/CD) pipelines, maintain deterministic behavior for reproducibility, and operate reliably without specialized deployment environments.[12–14]

We propose a middle path that leverages multiple LLM providers through structured workflows without requiring complex multi-agent infrastructure. We have named it Simple Lightweight Ensemble Analysis Network (SLEAN). We aim at a minimal-overhead Python setup that runs smoothly on a local machine. Our approach treats different providers as complementary analytical resources orchestrated through simple, deterministic protocols. Rather than implementing long-lived agent processes with persistent state, we employ a static orchestration model with three fixed phases: independent analysis, cross-critique, and final arbitration. Our coordination protocol follows three phases that preserve independence while enabling structured synthesis.

In Phase 1, each provider receives the same task description with the same corpus and produces an independent audit without seeing the output of its peer(s). Inputs may be segmented per provider to respect context limits, but the source material is identical. These audits are saved verbatim as artifacts, which fixes the baseline for later review and supports reproducible analysis. Providers never communicate directly; all interaction flows through the orchestrator with fixed prompts, which prevents early convergence and captures the provenance of every claim.

Phase 2 introduces cross-critique under symmetric prompts. Each provider reads the original inputs together with its own Phase 1 audit and the peer audit, then issues a consolidated assessment that records points of agreement, disagreements, and any proposed changes. These consolidations remain independent products, preserving diversity while exposing conflicts. In Phase 3 an arbitrating provider consumes the original inputs, both audits, and both consolidations and produces the definitive synthesis. The final artifact lists accepted and rejected fixes with justification and attribution, stores contribution percentages, and preserves every intermediate file in a versioned directory. This sequencing yields determinism at the workflow level, minimizes uncontrolled coupling between models, and provides an auditable trail that can be integrated into continuous delivery pipelines.

As a case study, we address a practical scenario common in software development, especially in this new vibe coding era: given a bug description and associated codebase, multiple independent LLM providers analyze the issue, critique each other's findings, and a final provider arbitrates the consensus fix. This workflow produces detailed analysis while maintaining the deployment simplicity required for production integration. It is light-weight and can easily run locally or without much cloud resources, making it practical and easily deployable.

**RELATED WORK**

There are several agent-oriented orchestrators. CrewAI implements role-based orchestration with hierarchical task delegation, AutoGen provides conversational workflows with dynamic group interactions, and LangGraph offers graph-structured coordination patterns.[8,15–17] Recent advances include ChatDev's virtual software teams with specialized communication protocols, MetaGPT's standardized operating procedures for structured coordination, and Swarm's scalable collaboration with memory sharing.[16,18,19] Domain-specific systems like FixAgent implement three-level coordination (Localizer, Repairer, Analyzer) while AGDebugger provides interactive debugging with user steering capabilities.[20,21] These systems require substantial infrastructure investment that limits practical adoption.[22,23]

A complementary foundation for coordinated analysis comes from ensemble learning, which shows that diversity across models can reduce variance and improve reliability when outputs are aggregated under a principled rule.[24,25] Classic work demonstrated that combining independent learners improves accuracy when their errors are decorrelated, a condition our Phase 1 enforces by isolating providers during independent audits.[26] Subsequent syntheses formalized voting, stacking, and bagging as general strategies for aggregation, which map naturally onto our deterministic arbitration in Phase 3.[24,27] Random Forests further quantified how ensemble error depends on member strength and inter-member correlation, reinforcing our emphasis on heterogeneity and controlled synthesis rather than unconstrained interaction.[28]

In automated bug investigation and repair, recent systems show that LLMs can localize faults without tests, suggesting that code and natural-language priors can substitute for traditional coverage signals in some settings.[29] Conversation-driven repair illustrates how multi-turn guidance improves patch quality on standard benchmarks, although it increases orchestration surfaces that must be engineered and audited in production.[30] Self-supervised repair pipelines use test-execution feedback to reduce human labelling while maintaining correctness, which underscores the value of reproducible control flow and persisted artifacts when models propose many candidate changes.[31] Generation-centric studies in multilingual settings show that disciplined selection and filtering of model outputs can be an effective baseline, aligning with our focus on consensus formation and explicit rejection of non-causal edits.[32]

Operationally, bringing learning systems into continuous delivery places a premium on transparency and auditability of decisions, which motivates our design choice to persist intermediate artifacts and attributions across phases.[33] Interpretable reporting frameworks advocate standardized summaries of model behavior and limitations to support downstream accountability, a practice our versioned outputs and contribution ledgers make concrete for bug investigation workflows.[34] Experience reports in software engineering for machine learning highlight the engineering frictions of integrating model-driven components into CI pipelines, which our configuration-driven orchestration addresses by fixing prompts, parameters, and file artifacts for reproducible runs.[35]

**KEY DIFFERENCES OF SLEAN**

Our contribution is a deterministic, file-driven coordination protocol across multiple LLM providers that yields auditable artifacts at each phase (independent audits, cross-consolidations, and arbitrated synthesis) and integrates cleanly with CI/CD. Unlike conversational multi-agent controllers that rely on long-lived agent state and role negotiation, our design enforces isolation in Phase 1, recorded critique in Phase 2, and a single, attributable arbitration in Phase 3, aligning with ensemble-learning principles for diversity and controlled aggregation while emphasizing reproducibility and provenance for production workflows. **Table 1** shows the key SLEAN differences.

*Table 1.* Differences between SLEAN and representative prior studies.

| AXIS | SLEAN | LITERATURE | WHY IT MATTERS FOR PRACTICE |
|---|---|---|---|
| EXECUTION MODEL | Deterministic orchestration of multiple LLM providers through a fixed three-phase workflow: independent analysis, cross-critique, and final arbitration. Static orchestration model with three strictly ordered phases that yields determinism at the workflow level. | Multi-agent frameworks require complex, stateful runtime environments with persistent agent memory, complex role negotiation, and graph-based orchestration (CrewAI hierarchical task delegation, AutoGen conversational workflow, LangGraph graph-structured coordination, ChatDev virtual software teams, and MetaGPT standardized operating procedures). [8,15,17,18,36] | Deterministic, repeatable runs reduce orchestration variance and enable precise regression analysis during CI/CD. [37–39] |
| STATE & ISOLATION | Providers never communicate directly; all interaction flows through the orchestrator with fixed prompts. Phase 1 enforces complete independence where providers do not see peer outputs. Provider independence enforced by orchestrator mediation. | Multi-agent systems maintain persistent agent state and enable direct communication between agents, with complex messaging protocols and role negotiation (CrewAI role-based orchestration, AutoGen dynamic group interactions, ChatDev specialized communication protocols, domain-specific systems like FixAgent three-level coordination). [8,15,18,21] | Early isolation reduces correlated errors and supports decorrelated evidence before aggregation, a classic ensemble prerequisite. [11,25,26] |
| AGGREGATION PRINCIPLE | Single arbitrating provider receives all prior artifacts and produces definitive synthesis with explicit accept/reject decisions and guidance. | Multi-agent frameworks rely on conversational consensus or graph search without deterministic, attributable combination rules (Graph-of-Thoughts collaborative reasoning, AutoAgents graph search, and ChatDev conversational consensus). [11,16,18] | Controlled aggregation maps to voting/bagging/stacking ideas and lowers variance when members are diverse while keeping a clear audit trail. [15,25,26,40,41] |
| PROVIDER STRATEGY | Treats providers such as OpenAI GPT and Anthropic Claude as interchangeable analytical resources via unified API abstraction. Exploits analytical diversity without role negotiation or provider-specific specializations. | Multi-agent systems typically instantiate roles over a single model family or within one vendor runtime (CrewAI single-model orchestration, AutoGen single-vendor focus, ChatDev homogeneous teams, MetaGPT standardized procedures). [8,15,17,18,36] | Cross-provider diversity increases hypothesis variety, which ensembles leverage for reliability when errors are partially independent. [25,26,36] |
| PROVENANCE & AUDITABILITY | Preserves complete artifacts for reproducible audits. Orchestrator constructs fresh, zero-padded results directory for every run. Every phase emits versioned files with attribution and rejection rationales persisted. | Multi-agent frameworks provide inconsistent artifact reporting and attribution tracking, with debugging challenges for complex agent interactions (human-factors work highlights debugging challenges, and model-reporting standards call for structured summaries). [21,42] | Provenance supports post-hoc accountability and reproducibility expectations in production ML & other software engineering pipelines. [35,43,44] |
| CI/CD INTEGRATION | File-driven, configuration-based design eliminates infrastructure overhead. Configuration parameters primarily defined in YAML files. Provides auditable trail that can be integrated into continuous delivery pipelines. | Multi-agent frameworks require specialized deployment environments with complex infrastructure dependencies and stateful services (substantial infrastructure investment limits adoption, and operational frictions occur when systems depend on stateful services). [13,23,45] | Pure file artifacts mesh with reproducible-builds ethos and supply-chain integrity practices, and predictable runs ease rollbacks and diff-based reviews. [12,46–48] |
| EXPLORATION VS. SCOPE CONTROL | Cross-review explores breadth through symmetric prompts; arbitration applies strict scope discipline, privileging minimal causal remedy over broader edits. | Multi-agent exploration through conversational reasoning can expand scope without explicit accept/reject accounting (conversational consensus in ChatDev, graph search in AutoAgent, role negotiation in CrewAI). [8,11,18] | Strict pruning limits blast-radius risk and review load, consistent with disciplined change-management in production systems. [49–51] |
| LINK TO THEORY & EVIDENCE | Explicitly aligns with ensemble-learning principles: Phase 1 isolation enforces decorrelated errors condition from classic ensemble theory; arbitration implements principled aggregation; outcomes fully attributable. | Multi-agent frameworks emphasize collaborative reasoning and role specialization without explicit grounding in ensemble learning principles for diversity and controlled aggregation (role-based specialization, conversational workflows, and virtual team coordination). [8,11,18] | The design operationalizes ensemble benefits while satisfying engineering requirements for traceability and reproducibility. [24,25] |

## METHODOLOGY AND SYSTEM DESIGN

**System Architecture Overview**

SLEAN is a layered multi-provider LLM framework that leverages complementary model strengths under a deterministic protocol and preserves complete artifacts for reproducible audits. Providers are treated as interchangeable analyzers behind a unified abstraction layer, enabling substitution or integration without infrastructure changes or role negotiation. The design eliminates direct provider-to-provider messaging to avoid the hidden state while still aggregating diverse reasoning outputs. An orchestrated three-phase analysis workflow maintains provider independence and enables systematic peer review and consensus formation.

**Figure 1** illustrates the simplified architecture and the controlled flow of information through five principal layers. The Input Layer manages all inbound resources, including configuration files, environment variables, bug descriptions, relevant codebase folders, and prompt templates. It also supports utility operations such as connection testing and initial project setup. Inputs are prepared into a structured form suitable for analysis before processing.

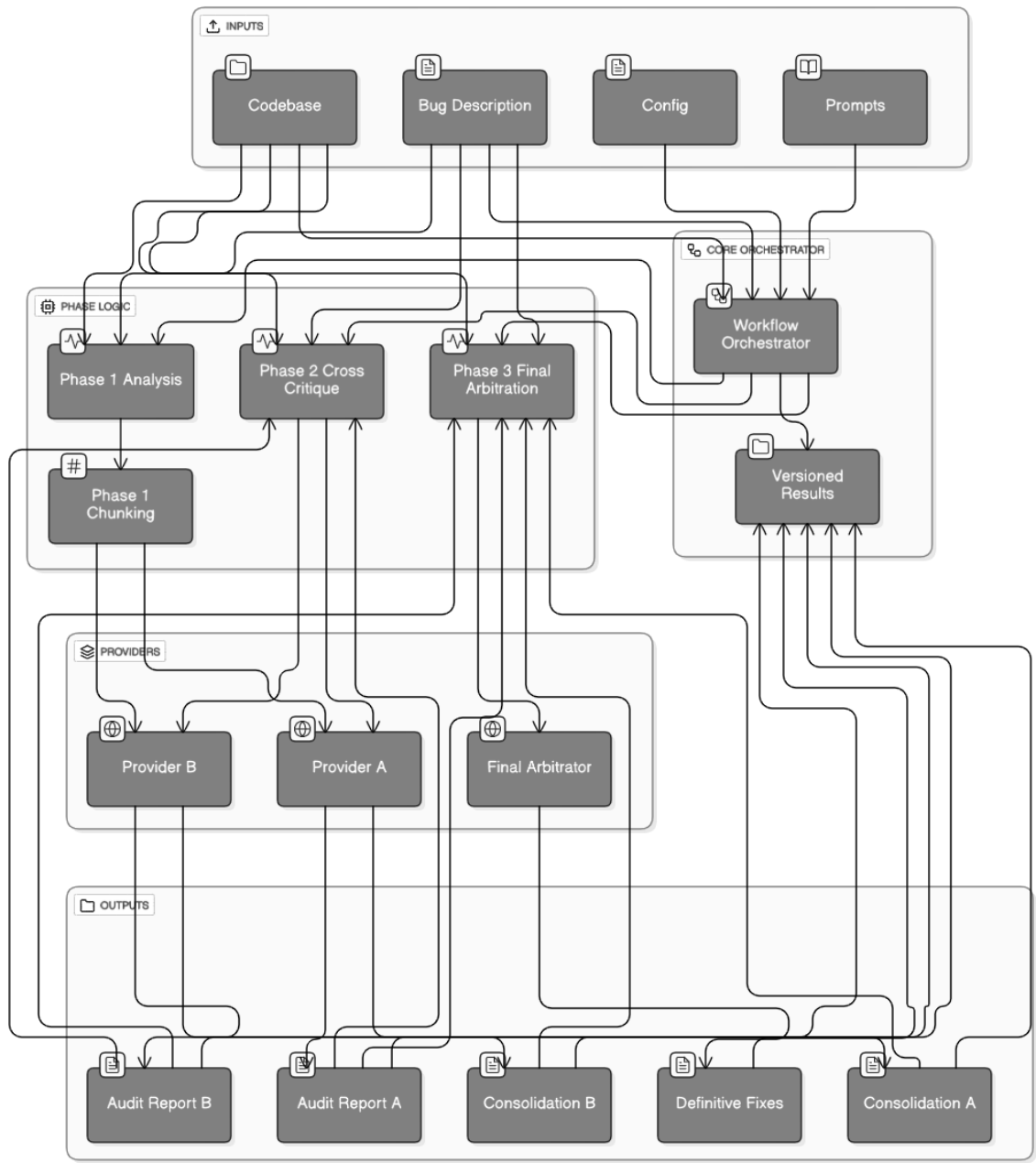

*Figure 1. SLEAN diagram showing its different components and phases.*

**Provider Abstraction and Coordination Framework**

The system uses four input categories at the start: Bug Description (problem statement with constraints), Codebase (recursive folder scan with supported extensions excluding build/cache directories), Configuration (YAML/JSON with environment variable substitution for ${VARNAME} patterns), and Prompts (phase-specific templates). The orchestrator creates zero-padded results directory (bug0001_results, bug0002_results) ensuring lexicographic ordering, copies input files for complete provenance, and rewrites all output paths to the versioned directory.

Configuration loading initializes dotenv early, performs environment substitution within APIs section, and instantiates providers through factory pattern behind the standardized AIClient interface. The abstract base class enforces explicit temperature and max_tokens parameters, where construction fails if either is missing, and provides uniform _make_api_request wrapper with exponential back-off and bounded retries, surfacing HTTP payloads for diagnosis. OpenAIClient normalizes endpoints to ensure exactly one /v1 segment before calling chat-completions; AnthropicClient uses x-api-key authentication with required anthropic-version header posting to /v1/messages.

Corpus ingestion performs a recursive folder scan collecting supported extensions in stable sorted order, reads files with UTF-8 decoding and recovery path for non-UTF-8 bytes, and assembles structured text with clear delimiters (# ===== filename =====) enabling safe chunking. Prompt management validates prompts folder existence and ensures referenced files are non-empty, externalizing content for domain-agnostic template swapping.

**Phase 1. Independent Analysis with Token Management**

The orchestrator routes bug description and structured codebase to Phase 1 Analysis, which applies token management before provider distribution. The token manager estimates tokens from character counts using provider-specific ratios, applies provider-specific context limits with 75% safety margin for prompt scaffolding and response generation. When content exceeds limits, it segments along file boundaries using structured headers from corpus ingestion, truncating single oversized files at line boundaries with explicit elision markers.

Each provider analyzes the same sources, namely a task description and an evidence corpus, subject to provider specific token limits that can require segmenting the corpus for each provider. Given by (eq. 1),

$$A_i = P_i(task_{description}, corpus\_chunks_i) \quad \text{(eq. 1)}$$

where $P_i$ is provider $i$ under the unified client interface, $task_{description}$ is the problem statement with any constraints, $corpus$ is the full evidence set such as code, documents, data, or logs, and $corpus\_chunks_i$ denotes the segments of that corpus prepared for provider $i$ to satisfy token limits. Providers do not see peer outputs in this phase. The analyses $A_i$ are saved as structured reports and establish the baseline. Providers never communicate directly; all routing occurs through orchestrator maintaining architectural isolation.

**Phase 2. Cross-Review and Consolidation.**

Each provider receives the task description, the full corpus, its own Phase 1 report, and the peer's Phase 1 report, and then produces a consolidated analysis. Prompts are symmetric in the sense that they distinguish own from other, rather than being literally identical. Given by (eq. 2),

$$C_i = P_i(task_{description}, corpus, A_i, A_{j \neq i}) \quad \text{(eq. 2)}$$

where $A_{j \neq i}$ is the Phase 1 analysis from the peer provider where $j \neq i$. This step enables peer review that can surface disagreements, omissions, or errors, while each $C_i$ remains the work of a single provider, which preserves analytical diversity.

**Phase 3. Synthesis and Arbitration.**

An arbitrating provider receives all prior artifacts together with the original inputs and produces the final output. Given by (eq. 3),

$$F = P_a(task_{description}, corpus, \bigcup_{i=1}^{n} A_i C_i) \quad \text{(eq. 3)}$$

where $P_a$ is the arbitrator selected in the workflow configuration and $F$ is the definitive result such as recommendations, decisions, or an action plan, and we have the collection of all analyses from multiple providers. Optional elements such as attribution or confidence follow the prompt template and are not enforced by program logic.

The sequential structure, explicit configuration that requires at least temperature and max tokens, and persisted artifacts support reproducible audits of runs. Independence in Phase 1 reduces early coupling, and peer review in Phase 2 supports systematic error detection and broader coverage. Full determinism or bias elimination is not guaranteed by code and depends on model settings and prompt design.

**Results Collection and Audit Trail**

All five outputs (Audit Report A, Audit Report B, Consolidation A, Consolidation B, Definitive Fixes) flow into the versioned results directory alongside copied input files. This creates a complete audit trail enabling reproducibility verification, run comparison, and systematic analysis of provider behavior across different configurations. Error handling flows uniformly through provider abstraction layer with consistent failure surfacing regardless of underlying provider implementation.

The sequential structure with explicit configuration requirements, provider isolation through orchestrated routing, and comprehensive data persistence supports reproducible audits while accommodating diverse provider capabilities within controlled execution envelope.

*Table 2.* Bug types and general information.

| Bug ID | Domain | Number of Files | Bug Description (chars) | Codebase (Chars) |
|---|---|---|---|---|
| 001 | Authentication | 4 | 518 | 4395 |
| 002 | Algorithmic Error | 4 | 84 | 8332 |
| 003 | Logic Error | 4 | 107 | 10891 |
| 004 | Transaction Integrity | 4 | 2607 | 11453 |
| 005 | Data Synchronization | 5 | 194 | 16647 |
| 006 | Null Reference | 4 | 1194 | 17793 |
| 007 | Distributed Systems | 6 | 64 | 35211 |
| 008 | Configuration | 5 | 1225 | 3266 |
| 009 | Concurrency | 7 | 353 | 5773 |
| 010 | Regex Performance | 5 | 1213 | 11096 |
| 011 | Race Condition | 5 | 7006 | 13906 |
| 012 | Server Imbalance | 4 | 760 | 21109 |
| 013 | Wrong Cache and Calculations | 5 | 92 | 26699 |
| 014 | Scheduling and Parsing | 5 | 145 | 13649 |
| 015 | Data Validation/Edge Case | 6 | 164 | 40499 |
| Average | - | ~5 | ~1,048 | ~16,048 |

## CASE STUDY METHODS

### Data Sources and Collection

We systematically analyzed 15 software debugging scenarios, each consisting of a deliberately buggy codebase paired with a casual user bug report. Each scenario was documented through structured definitive fix reports (definitive_fixes_001.md through definitive_fixes_015.md) containing AI-assisted debugging performance metrics, contribution assessments, proposed solutions, and final arbitration decisions.

The experimental design employed autogenerated codebases with strategically embedded bugs to simulate realistic debugging challenges, especially found by vibe coders. Codebases were intentionally provided without bug identification markers, and diagnostic information was deliberately limited to increase discovery difficulty for AI systems. Bug selection criteria prioritized single-run resolvable issues, while some codebases contained multiple bugs, non-crashing bugs were co-located to enable simultaneous resolution, whereas crash-inducing bugs were isolated as single defects to prevent multi-stage debugging requirements.

The dataset encompasses diverse codebase complexities (**Table 2**) and bug report verbosity levels, ranging from minimal 64-character casual descriptions to comprehensive 7,006-character diagnostic logs. The codebase sizes varied from compact 3,266-character implementations to complex 40,499-character systems, enabling evaluation across different code comprehension and analysis demands.

### User Input Scenario Classification

User prompts were extracted from bug report files (bug_001.txt through bug_015.txt) and systematically categorized based on information completeness and diagnostic complexity. To simulate realistic debugging scenarios encountered by AI systems, we deliberately employed casual, non-technical language typical of general public users rather than professional developers ("vibe coder" simulation). Stack traces, where present, consisted of authentic terminal output copied directly from error logs without modification.

### Configurations

Experiments used a Python runner (main.py) with YAML configuration managing provider assignments, model specifications, and parameters. AI[A] was assigned to OpenAI (gpt-3.5-turbo, temperature=0.1, max_tokens=3000), AI[B] to Anthropic (claude-sonnet-4-20250514, temperature=0.1, max_tokens=4000), and the final arbitrator to OpenAI with identical AI[A] settings. This created an OpenAI-Anthropic-OpenAI workflow with retry logic (3 attempts, exponential backoff) and API authentication via environment variables.

Pre-run validation verified API connectivity and token limits using character-to-token ratios of 0.25, with context limits at 16,000 tokens (OpenAI) and 200,000 tokens (Anthropic), applying 75% safety margins before automatic chunking (applied only in Phase 1 for initial testing; see Future Work). The workflow consumed four inputs: bug.txt (problem description and traces), codebase/ directory (source code), config.yaml (environment variables), and three prompt templates. Three distinct prompt templates controlled each workflow phase. The bug_slayer_prompt.txt directed Phase 1 independent analysis, requiring structured output with executive summary, root cause analysis, prioritized fixes with exact code snippets, implementation strategy, and testing recommendations. audit_consolidator_prompt.txt guided Phase 2 cross-critique through systematic methodology: technical accuracy assessment, investigation depth evaluation, fix quality analysis, and evidence strength verification, culminating in superior synthesis with quantified contribution percentages. final_consolidator_prompt.txt enforced Phase 3 arbitration via contradiction detection, peer review validation, and five quality gates (technically sound, peer validated, non-conflicting, minimal impact, evidence-backed) with explicit rejection criteria and directness requirements.

Source code was recursively scanned excluding build directories, filtering 24 file extensions (.py, .js, .java, .cpp, .c, .h, .cs, .php, .rb, .go, .rs, .ts, .jsx, .tsx, .vue, .swift, .kt, .scala, .r, .sql, .html, .css, .json, .xml, .yaml, .yml, .md, .txt), and assembled with UTF-8 encoding into structured text with file boundary delimiters and metadata headers.

Each run executed a deterministic sequence: Phase 1 to 3 as identical prompts (bug description + codebase) sent to $AI^A$ and $AI^B$ independently for parallel analysis. Phase 2 enabled cross-critique where each AI reviewed the peer's Phase 1 output alongside original inputs. Phase 3 provided all previous artifacts to the arbitrator for final synthesis. Each run generated versioned directories (bug0001_results, etc.) containing input copies and artifacts: audit_report_A.md ($AI^A$ analysis), audit_report_B.md ($AI^B$ analysis), consolidation_A.md ($AI^A$ cross-critique), consolidation_B.md ($AI^B$ cross-critique), and definitive_fixes.md (arbitrator synthesis).

AI contribution percentages were extracted directly from standardized headers in each definitive fix document. Four contribution categories were systematically recorded: (i) $AI^A$ CONTRIBUTION, representing the first AI system's individual contribution; (ii) $AI^B$ CONTRIBUTION, representing the second AI system's individual contribution; (iii) CONVERGENT (BOTH AGREED), representing collaborative solutions where both AI systems reached consensus; and (iv) FINAL ARBITRATION, representing the arbitrator's intervention, or final reviewer, which in this case is also $AI^B$.

We employed systematic content analysis to quantify debugging solutions. Accepted fixes were identified by enumerating all items listed under "DEFINITIVE FIX LIST" sections, with each Fix #N entry counted as one accepted proposition. Rejected fixes were quantified by cataloguing all items explicitly listed in "REJECTED FIXES" sections of each document. Total fix propositions were calculated as the sum of accepted and rejected fixes. This approach ensured comprehensive capture of all debugging attempts while maintaining clear distinction between successful and unsuccessful solutions.

Acceptance rates were computed using the formula: Acceptance Rate = (Accepted Fixes / Total Fix Propositions) × 100, with results expressed as percentages rounded to one decimal place. This metric provided a quantitative measure of debugging efficiency for each scenario.

Several composite metrics were calculated from primary data. Total AI contribution was computed as $AI^A + AI^B$, representing the combined two AIs without the third AI arbitration. Max Contributor was determined by identifying the category ($AI^A$, $AI^B$, Convergent, or Arbitrator) with the highest percentage contribution for each bug scenario.

**Analysis**

Two binary classification schemes were implemented. The Arbitration Flag was set to TRUE when Final Arbitration ≥ 50%, indicating scenarios requiring substantial extra (arbitrator layer) intervention. The Convergence Flag was set to TRUE when Convergent contribution ≥ 75%, identifying cases where AI systems achieved high levels of agreement without the arbitration layer.

Data extraction underwent systematic verification through document-by-document audit procedures. Each contribution percentage was cross-referenced against source document headers using exact text matching. Fix counts were validated by manual enumeration of all Fix #N entries and rejected approach listings. Calculated fields were verified through independent mathematical computation. Boolean flag assignments were confirmed through rule-based logic verification. This multi-stage validation process ensured 100% data integrity across all table entries. The bug fixes were checked manually, applied and tested in the code, ensuring that they fixed the proposed bugs.

The analysis maintained strict adherence to source document specifications without statistical interpolation or estimation. All percentages, counts, and derived metrics were computed deterministically from explicitly stated values in the definitive fix documentation. No data imputation or statistical modeling was employed, ensuring direct traceability between table entries and source evidence.

All source documents (definitive_fixes_001.md through definitive_fixes_015.md) and extraction protocols are available for independent verification and replication of results at github.com/cyaura-open.

All fixes were independently validated by applying each fix to the buggy codebase, executing the program with test inputs, and verifying resolution of the reported behavior. Rejected fixes were reviewed to confirm they would introduce errors, break functionality, or expand scope unnecessarily.

**RESULTS**

We identified two distinct input categories (**Table 3**) based on the availability of technical diagnostic information: (i) Technical/Diagnostic Issues (n=8, 53%), providing substantial technical information including complete terminal error output with file locations and exception types, detailed system logs, reproduction commands, or comprehensive execution traces, paired with casual problem descriptions; and (ii) Functional/Behavioral Issues (n=7, 47%), containing informal problem statements using everyday language without technical diagnostics, describing behavioral inconsistencies or system malfunctions through user-observable symptoms.

*Table 3. Bug description by the users divided into 2 categories.*

| CATEGORY | DEFINITION | BUG IDS | N | % |
|---|---|---|---|---|
| TECHNICAL/DIAGNOSTIC ISSUES (TYPE 1) | Substantial technical information including stack traces, detailed system output, reproduction commands, or comprehensive execution logs | 001, 004, 006, 008, 009, 010, 011, 012 | 8 | 53 |
| FUNCTIONAL/BEHAVIORAL ISSUES (TYPE 2) | Informal problem statements using everyday language without technical diagnostics, describing behavioral inconsistencies or user-observable symptoms | 002, 003, 005, 007, 013, 014, 015 | 7 | 47 |

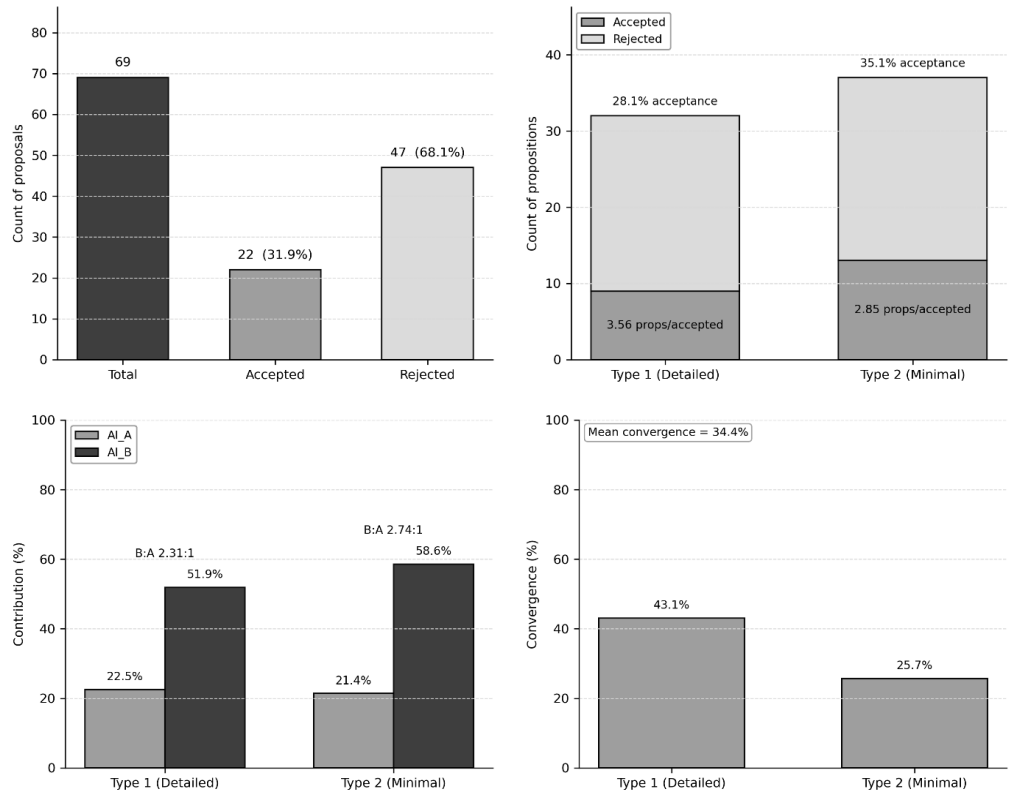

**Figure 2.** (top-left) Distribution of total, accepted, and rejected bug fix propositions. (top-right) Fix acceptance rates and efficiency by input description type. (bottom-left) Relative AI contribution for the fixes by input type. (bottom-right) Convergence rates between AI systems by input type

Technical/Diagnostic Issues included: Bug 001: "User registration fails with 500 error" [authentic TypeError stack trace]; Bug 004: "something is clearly incorrect in my system, fix it" [complete banking transaction terminal output]; Bug 006: "the code crashed" [authentic AttributeError stack trace]; Bug 008: "code crashes" [authentic ImportError circular import stack trace]; Bug 009: "I have a nasty bug once I run the below in powershell" [PowerShell load testing command]; Bug 010: "bugs when I run the below" [curl command with full HTTP response output]; Bug 011: "my results seem incorrect, there is a hidden bug in my code" [extensive job scheduler execution logs]; Bug 012: "something is odd with the load distribution of servers" [detailed load distribution metrics and capacity analysis].

Functional/Behavioral Issues included: Bug 002: "my calculations seem to be incorrect. fix them." [no technical details]; Bug 003: "something is wrong with my management system but I am unsure what it is" [vague description]; Bug 005: "Different ways of counting tasks in our system give different results" [behavioral inconsistency]; Bug 007: "Our cache system is completely broken!" [emotional, non-technical language]; Bug 013: "different ways of calculating customer revenue give different results" [calculation discrepancy]; Bug 014: "it is failing to scheduling there is a problem with the slots and there is a isoformat string that is invalid" [scheduling malfunction]; Bug 015: "The system is still counting stocks I sold as if I still own them, which makes my portfolio value calculations completely wrong" [portfolio calculation error].

SLEAN was tested across 15 software bugs spanning diverse fault domains, with system-wide performance metrics for both bug input types presented in **Figure 2**. The complete dataset comprised 69 individual fix propositions, of which 22 were accepted and 47 were rejected, yielding an overall acceptance rate of 31.9% (95% CI: 20.9–42.9%).

We stratified the dataset by input format: Type 1 inputs (detailed diagnostic information, n = 8) versus Type 2 inputs (minimal colloquial descriptions, n = 7). Type 1 inputs generated 32 total propositions with 9 acceptances, for a 28.1% acceptance rate.

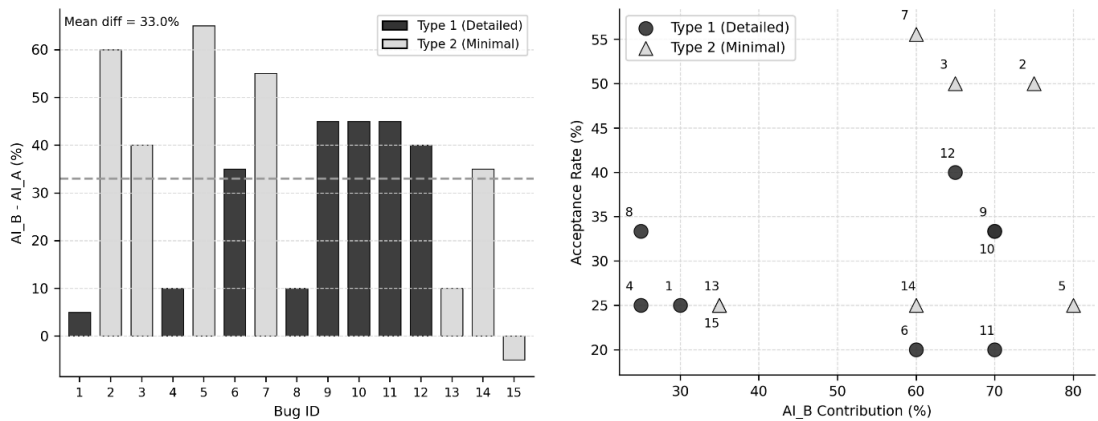

*Figure 3. (left)* $AI^B$ dominance over $AI^A$ in contributions when fixing specific bugs. *(right)* Acceptance rate according to the bug input type and $AI^B$ contribution.

Type 2 inputs produced 37 propositions with 13 acceptances or a 35.1% acceptance rate. Type 2 inputs demonstrated superior efficiency, requiring 2.85 propositions per accepted fix compared to 3.56 for Type 1 inputs, representing a 20% improvement in resource utilization.

The AI systems exhibited markedly different response patterns across input types. For Type 1 inputs, $AI^A$ contributed a mean of 22.5% of solutions while $AI^B$ contributed 51.9%. Type 2 inputs showed even greater asymmetry, with $AI^A$ contributing 21.4% and $AI^B$ contributing 58.6%. This produced $AI^B$-to-$AI^A$ contribution ratios of 2.31:1 for Type 1 inputs and 2.74:1 for Type 2 inputs. Convergence rates varied inversely, with Type 1 inputs achieving higher mean convergence (43.1%) compared to Type 2 inputs (25.7%).

Across all bugs, $AI^B$ demonstrated consistent dominance in solution generation (**Figure 3**). $AI^B$ outperformed $AI^A$ by an average of 33.0%, with $AI^B$ exceeding $AI^A$ contributions in 14 of 15 cases (93.3%). Only bug 015 (Data Validation/Edge Case domain) showed $AI^A$ dominance, where $AI^A$ contributed 40% versus $AI^B$'s 35%. The largest performance gaps occurred in bugs 002 (Algorithmic Error, +60%) and 005 (Data Synchronization, +65%), while the smallest differences appeared in bugs 001 (Authentication, +5%) and 004 (Transaction Integrity, +10%).

Total AI contribution (combined $AI^A$ and $AI^B$ percentages) averaged 77.0% across all bugs, with the arbitrator contributing the remaining 23.0%. The distribution of total AI contribution varied substantially, ranging from 40% in bugs 004 and 008 to 95% in bugs 005, 009, 010, and 011, indicating significant variation in the complexity of arbitrator oversight required across different bug types.

In contrast, six domains showed consistently modest performance. Authentication, Transaction Integrity, Data Synchronization, Wrong Cache and Calculations, Scheduling and Parsing, and Data Validation/Edge Case domains all achieved exactly 25% acceptance rates. Configuration defects achieved 33.3% acceptance but required arbitration. Concurrency-related bugs (Concurrency, Regex Performance, Race Condition) demonstrated high convergence rates (85-90%) but modest acceptance rates ranging from 20% - 33.3%.

Codebase complexity metrics showed no systematic relationship with debugging success. The number of files ranged from 4 to 7 across all bugs, while bug description length varied dramatically from 64 characters (Distributed Systems) to 7,006 characters (Race Condition), with no apparent correlation to acceptance outcomes. Similarly, codebase size varied from 3,266 characters to 40,499 characters without systematic impact on fix acceptance rates.

AI system convergence showed minimal relationship with ultimate fix acceptance (**Figure 4. left**), with high convergence (≥80%) occurring in 4 of 15 cases: Logic Error (85%), Concurrency (90%), Regex Performance (85%), and Race Condition (85%). These high-convergence cases averaged 34.1% acceptance compared to 31.7% for all other cases, a difference of only 2.4 percentage points. However, convergence did not serve as a perfect predictor of arbitration necessity: both arbitrated cases (bugs 004: Transaction Integrity and 008: Configuration) exhibited convergence rates of exactly 10%, while several non-Arbitrated bugs also had 10% convergence (e.g., 002, 005, 007, 012), and no bugs with convergence above 10% required arbitration. The four high-convergence cases involved technically well-defined domains with clear behavioral specifications, with two of these four cases representing concurrency-related bugs (bugs 009 and 011), indicating particular AI system alignment on timing-dependent fault patterns. In arbitrated cases, the arbitrator AI contributed 50% of the final solutions, representing substantial intervention.

*Table 4. General table of studied bugs.*

| Bug | Domain | Propositions | Acceptance Rate (%) | Codebase Files | AI[A] (%) | AI[B] (%) | Arbitrator (%) |
|---|---|---|---|---|---|---|---|
| 1 | Authentication | 4.0 | 25.0 | 4 | 25.0 | 30.0 | 5.0 |
| 2 | Algorithmic Error | 2.0 | 50.0 | 4 | 15.0 | 75.0 | 0.0 |
| 3 | Logic Error | 2.0 | 50.0 | 4 | 25.0 | 65.0 | 10.0 |
| 4 | Transaction Integrity | 4.0 | 25.0 | 4 | 15.0 | 25.0 | 50.0 |
| 5 | Data Synchronization | 4.0 | 25.0 | 5 | 15.0 | 80.0 | 5.0 |
| 6 | Null Reference | 5.0 | 20.0 | 4 | 25.0 | 60.0 | 0.0 |
| 7 | Distributed Systems | 9.0 | 55.6 | 6 | 5.0 | 60.0 | 25.0 |
| 8 | Configuration | 3.0 | 33.3 | 5 | 15.0 | 25.0 | 50.0 |
| 9 | Concurrency | 3.0 | 33.3 | 7 | 25.0 | 70.0 | 5.0 |
| 10 | Regex Performance | 3.0 | 33.3 | 5 | 25.0 | 70.0 | 5.0 |
| 11 | Race Condition | 5.0 | 20.0 | 5 | 25.0 | 70.0 | 5.0 |
| 12 | Server Imbalance | 5.0 | 40.0 | 4 | 25.0 | 65.0 | 0.0 |
| 13 | Wrong Cache and Calculations | 8.0 | 25.0 | 5 | 25.0 | 35.0 | 10.0 |
| 14 | Scheduling and Parsing | 8.0 | 25.0 | 5 | 25.0 | 60.0 | 0.0 |
| 15 | Data Validation/Edge Case | 4.0 | 25.0 | 6 | 40.0 | 35.0 | 5.0 |
| - | **Average** | **4.6** | **32.4** | **4.9** | **22.0** | **55.3** | **11.7** |

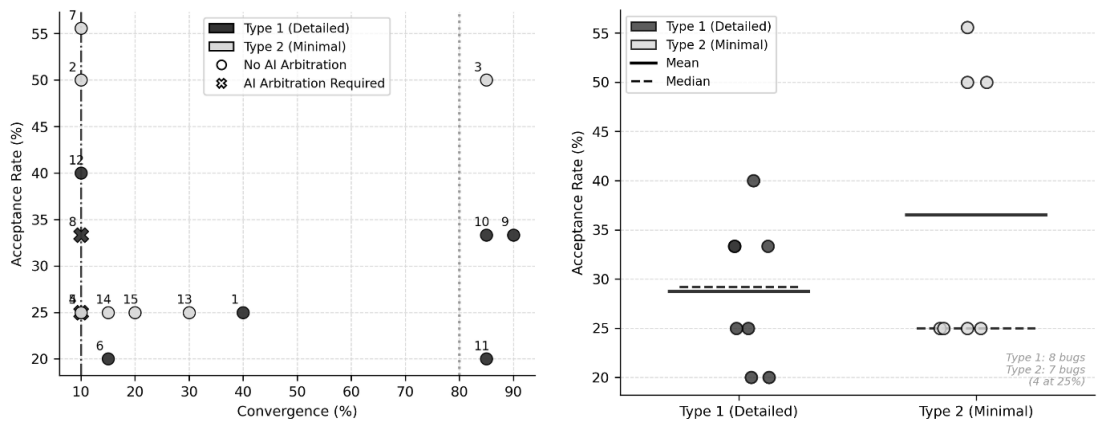

**Figure 4. (left)** Acceptance rate of bug fixes, convergence between AI[A] and AI[B], and arbitration flag at 80%. **(right)** Distribution of acceptance rate according to bug input type.

The relationship between proposal quantity and acceptance success revealed complex efficiency patterns strongly influenced by input format as seen in **Figure 4 (right)**. Type 1 (detailed) inputs clustered consistently in moderate efficiency ranges (3–5 propositions, 20–40% acceptance), suggesting that detailed diagnostic information provides sufficient context to avoid major confusion while potentially introducing noise that prevents highly efficient solutions. By contrast, Type 2 (minimal) inputs exhibited extreme outcomes (25–55.6% acceptance), either achieving the highest efficiency (2 propositions, 50% acceptance for bugs 002: Algorithmic Error and 003: Logic Error) or requiring extensive exploration (8–9 propositions). **Table 4** shows all the data summarized.

This "input minimalism paradox" demonstrates that sparse problem descriptions either provide sufficient clarity for rapid AI convergence or prove so ambiguous that multiple solution iterations become necessary (**Table 5**). Bug 007 (Distributed Systems) exemplifies this pattern, requiring 9 propositions yet achieving the highest acceptance rate (55.6%), a pattern consistent with architectural issues benefiting from multiple targeted interventions, though this dataset does not establish causality. Excluding this architectural outlier, high-volume bugs (8 propositions, bugs 013 and 014) achieved only 25% acceptance rates.

The distribution of maximum contributors revealed clear system specialization patterns: AI[B] served as the maximum contributor in 7 cases, convergent agreement occurred in 5 cases, the arbitrator ("Final") dominated in 2 cases, and AI[A] led in only 1 case. This distribution reinforces the systematic advantage of AI[B] across diverse fault types, confirms the rarity of scenarios where AI[A]-centric approaches prove superior, and demonstrates substantial AI system alignment on technically well-defined problems.

*Table 5. Debugging efficiency patterns by proposal volume and input format.*

| Propositions | Bugs (Input Type) | Mean Acceptance | Input Type Pattern | Insight |
|---|---|---|---|---|
| 2 | 2(T2), 3(T2) | 50.0% | All Type 2 | Highest efficiency |
| 3 | 8(T1), 9(T1), 10(T1) | 33.3% | All Type 1 | Consistent moderate |
| 4 | 1(T1), 4(T1), 5(T2), 15(T2) | 25.0% | Mixed (2T1, 2T2) | Lower efficiency |
| 5 | 6(T1), 11(T1), 12(T1) | 26.7% | All Type 1 | Mixed Type 1 results |
| 8 | 13(T2), 14(T2) | 25.0% | All Type 2 | Low efficiency |

*Table 6 - Risk categories for AI-generated code fixes with operational impacts and arbitration decision patterns.*

| Risk Category | Representative Examples from Dataset | Documented Operational Risk | Arbitration Decision Pattern |
|---|---|---|---|
| Syntax Errors and Technical Impossibilities | AI[A]'s fatal syntax error with threading.Lock() without storing the instance (Bug 009); AI[A]'s invalid threading syntax causing compilation errors (Bug 009) | Prevents code execution entirely, leading to immediate deployment failures and compilation errors in production | Rejected as "fatal syntax error that would prevent code from running." Basic technical correctness is required |
| Fixing Nonexistent Problems | AI[B]'s enhanced error handling where the original code already had proper bounds checking (Bug 002); AI[A]'s queue clearing fix when queue.clear() already works (Bug 007); AI[B]'s UUID solution as overly complex for a nonexistent collision problem (Bug 004) | Wastes development resources, introduces unnecessary code complexity, and obscures real issues by adding redundant layers | Rejected as "addresses theoretical problem not evidenced" or "safe function already exists and doesn't need modification" |
| Over-engineering and Unnecessary Complexity | AI[B]'s conversion to Decimal arithmetic without evidence of floating-point errors (Bug 002); AI[B]'s arbitrary 70/30 weighting with no proven benefit (Bug 012); multi-step approaches where the defect required a one-line fix (Bug 010) | Increases maintenance burden, complicates testing, and may reduce performance. Complex solutions are harder to debug and validate | Rejected as "unnecessary complexity," "speculative addition not required," or "overengineered when a simple fix suffices" |
| Breaking Existing Functionality | AI[B]'s hardcoding fix that removed environment-aware behavior (Bug 008); AI[A]'s resource tracking leading to leaks (Bug 011); AI[A]'s fix that could overload smaller servers (Bug 012) | Causes immediate service degradation, resource exhaustion, or loss of critical system capabilities; disrupts established operational patterns | Rejected as "breaks existing error handling," "would prevent legitimate completions," or "destroys functionality" |
| Misunderstanding Existing Code | AI[A]'s duplicate checking fix when identical code already existed (Bug 004); AI[A]'s consolidation that misidentified the root cause (Bug 005); AI[A]'s invalid import change (Bug 008) | Consumes development effort without resolving defects, reflects poor comprehension of system logic, and risks introducing further errors | Rejected as "misreading existing code," "demonstrates misunderstanding of the problem," or "failed to recognize existing implementations" |
| Addressing Symptoms Rather Than Root Cause | AI[A]'s fixes targeting the reporting module even though it was functioning correctly (Bug 005); AI[A]'s rounding adjustment addressing symptoms instead of cause (Bug 013); AI[B]'s defensive programming not tied to root cause (Bug 005) | Leaves actual defects unresolved, allowing recurrence under different conditions; increases code complexity while masking core issues | Rejected as "addresses symptoms, not root cause," "defensive programming without addressing cause," or targeting "functioning modules" |
| Production Code Quality Violations | AI[A]'s insertion of print statements into production code (Bug 007); AI[B]'s runtime import that degrades performance and violates clean code standards (Bug 012) | Reduces production efficiency, violates coding standards, and risks information leakage through improper logging | Rejected as "not a valid fix in production" or "creates performance penalty" |
| Speculative Enhancements Beyond Scope | Security measures without concrete implementation (Bug 001); documentation-focused changes instead of functional fixes (Bug 003); AI[B]'s unnecessary code removal (Bug 010) | Expands scope beyond defect resolution, delays critical fixes, and introduces untested or non-essential changes | Rejected as "speculative enhancement not required," "not necessary for immediate bug resolution," or "beyond scope" |

## DISCUSSION ON RISK-BEARING SUGGESTIONS AND FILTRATION OUTCOMES

The consolidations consistently surfaced extensive "improvement" suggestions that extended far beyond minimal defect resolution. Individual AI systems demonstrated systematic bias toward comprehensive refactoring, security hardening, and architectural modifications that would introduce substantial operational risk in production hot-fix scenarios. Final arbitration systematically pruned these broader edits as out-of-scope, preserving only minimal causal interventions. **Table 6** summarizes actual categories observed across fault domains, documenting where such suggestions originated, their documented operational risks, and arbitration disposition patterns.

## LIMITATIONS AND FUTURE WORK

Several architectural extensions could enhance SLEAN's capabilities. Sequential processing through phases prevents parallelization that could reduce execution time, particularly for independent analysis where providers could operate concurrently. Static configuration requires provider assignment at initialization, preventing dynamic adaptation based on availability or performance characteristics. File boundary chunking may struggle with extremely large individual files that exceed token limits even after partitioning.

The framework's modular architecture and clear abstraction boundaries explicitly support future extensions. The provider interface can accommodate new providers with minimal modification, while the prompt template system enables domain-specific customization. The workflow orchestration pattern supports extension with additional phases or modified coordination protocols without affecting core functionality.

The framework's modular design supports investigation of multilayered review architectures with expanded arbitration hierarchies. Future studies should examine n-layer workflows incorporating diverse arbitrators, heterogeneous provider combinations, and varied prompt engineering strategies. Information accumulation across layers where each phase preserves and augments prior artifacts represents a promising direction for comprehensive analytical depth.

Additional enhancements include parallel execution for independent analysis phases, adaptive provider selection based on bug characteristics or performance history, feedback integration mechanisms learning from successful repair sessions, and hierarchical analysis approaches for very large codebases decomposing complex systems into manageable components while maintaining overall coherence.

Another limitation of our experimental design is that the arbitrator shares the same underlying architecture as AI[A]. This may introduce bias during arbitration. Future work should employ an independent third provider for arbitration to ensure complete architectural diversity.

Chunking is currently implemented only in Phase 1. Tested codebases (3,266 to 40,499 characters) passed all phases without token violations, but larger repositories may require multi-phase chunking strategies. Future work will evaluate SLEAN on enterprise-scale codebases, implement parallel execution for independent analysis phases, and explore dynamic provider selection based on performance characteristics.

## CONCLUSION

We presented SLEAN, a practical framework for coordinating multiple LLM providers in automated bug investigation that achieves the analytical diversity of multi-agent systems without their infrastructure complexity. By treating each provider as an interchangeable analytical module within a deterministic three-phase workflow, the framework preserves independence, supports reproducible audits, and filters harmful proposals before code integration.

Evaluation demonstrated that the framework effectively identifies and rejects scope-expanding modifications while preserving minimal causal interventions. The systematic arbitration process proved essential in preventing potentially harmful AI suggestions from reaching production code, addressing a critical safety concern in AI-assisted development.

The approach generalizes beyond debugging to any domain requiring synthesis from heterogeneous LLM analyses. The provider-agnostic architecture and file-driven design make it readily applicable to security auditing, code review, document verification, and risk assessment. By combining operational simplicity, deterministic coordination, and explicit traceability, SLEAN offers a viable production-ready alternative to complex multi-agent systems where reliable, explainable, and auditable analytical consensus is required.

## CONFLICT OF INTEREST

No conflict of interest.

## EXTRA MATERIAL

https://github.com/cyaura-open/SLEAN

## ACKNOWLEDGEMENTS

Dr. Yong Je Kwon and Dr. Anany Dwivedi for reviewing this work.